\documentclass[12pt]{iopart}
\usepackage{iopams}  
\usepackage{caption}
\usepackage{color}
\usepackage{graphicx}
\begin{document}

\title  [Angular resolution \ldots GRAPES-3]   
        {The angular resolution of GRAPES-3 EAS array after correction for the shower front curvature}    

\author{
        V.B.~Jhansi$^{a,b}$,
        S.~Ahmad$^{a,c}$,
        M.~Chakraborty$^{a,b}$,
        S.R.~Dugad$^{a,b}$,
        S.K.~Gupta$^{a,b}$,
        B.~Hariharan$^{a,b}$,
        Y.~Hayashi$^{a,d}$,
        P.~Jagadeesan$^{a,b}$,
        A.~Jain$^{a,b}$,
        P.~Jain$^{a,e}$,
        S.~Kawakami$^{a,d}$,
        H.~Kojima$^{a,f}$,
        S.~Mahapatra$^{a,g}$,
	P.K.~Mohanty$^{a,b}$
        S.D.~Morris$^{a,b}$,
        P.K.~Nayak$^{a,b}$,
        A.~Oshima$^{a,f}$,
        D.~Pattanaik$^{a,b}$,
        P.S.~Rakshe$^{a,b}$,
        K.~Ramesh$^{a,b}$,
        B.S.~Rao$^{a,b}$,
        L.V.~Reddy$^{a,b}$,
        S.~Shibata$^{a,f}$,
        F.~Varsi$^{a,e}$,
        M.~Zuberi$^{a,b}$
        }

\address{$^a$   The GRAPES-3 Experiment, Cosmic Ray Laboratory, 
                Raj Bhavan, Ooty 643001, India}
\address{$^b$   Tata Institute of Fundamental Research, 
                Dr. Homi Bhabha Road, Mumbai 400005, India}
\address{$^c$   Aligarh Muslim University, 
                Aligarh 202002, India}
\address{$^d$   Graduate School of Science, 
                Osaka City University, Osaka 558-8585, Japan}
\address{$^e$   Indian Institute of Technology Kanpur, 
                Kanpur 208016, India}
\address{$^f$   College of Engineering, Chubu University, 
                Kasugai, Aichi 487-8501, Japan}
\address{$^g$   Utkal University, 
                Bhubaneswar 751004, India}

\eads{\mailto{pkm@tifr.res.in}}

\begin{abstract}
The angular resolution of an extensive air shower (EAS) array
plays a critical role in determining its sensitivity for the
detection of point $\gamma$-ray sources in the multi-TeV
energy range. The GRAPES-3 an EAS array located at Ooty in
India (11.4$^{\circ}$N, 76.7$^{\circ}$E, 2200\,m altitude) is
designed to study $\gamma$-rays in the TeV-PeV energy range.
It comprises of a dense array of 400 plastic scintillators
deployed over an area of 25000\,m$^2$ and a large area
(560\,m$^2$) muon telescope. A new statistical method allowed
real time determination of the propagation delay of each
detector in the GRAPES-3 array. The shape of shower front is
known to be curved and here the details of a new method 
developed for accurate measurement of the shower front
curvature is presented. These two developments have led to a
sizable improvement in the angular resolution of GRAPES-3
array. It is shown that the curvature depends on the size and
age of an EAS. By employing two different techniques, namely,
the odd-even and the left-right methods, independent estimates
of the angular resolution are obtained. The odd-even method
estimates the best achievable resolution of the array. For
obtaining the angular resolution, the left-right method is
used after implementing the size and age dependent curvature
corrections. A comparison of the angular resolution as a
function of EAS energy by these two methods shows them be
virtually indistinguishable. The angular resolution of
GRAPES-3 array is 47$^{\prime}$ for energies E$>$5\,TeV and
improves to 17$^{\prime}$ at E$>$100\,TeV and finally
approaching 10$^{\prime}$ at E$>$500\,TeV.
\end{abstract}
\maketitle

\section{Introduction} \label{sec:intro}
The origin of primary cosmic rays (PCRs) is a fundamental problem
of high-energy astrophysics which has remained unresolved even
after their discovery by Hess more than a century ago. The charged
nature of PCRs causes their path to be continuously modified due
to passage through space permeated by randomly oriented magnetic
fields. Consequently, their directions get completely randomised
that bear no relation to the sources of their origin. On the other
hand, the small flux of primary $\gamma$-rays, produced along with
the charged PCRs travel in a straight line from the sources of
their origin allowing detection of the sources of PCRs
\cite{Gaisser16}. Past two decades have witnessed phenomenal
progress in the field of $\gamma$-ray astronomy in the MeV-GeV
energy range and a very large number of sources have been
discovered by space-borne telescopes. However, in the GeV-TeV
energy range the $\gamma$-ray flux is too low to be detected by
the present modest area telescopes flown on the satellites. The
ground-based imaging atmospheric Cherenkov telescopes (IACTs)
which detect the faint Cherenkov light emitted by the electrons
in an extensive air shower (EAS) are ideally suited for the
detection of sub-TeV $\gamma$-rays \cite{Funk15}. The EAS arrays
enjoy a unique advantage due to the following two factors, (i) a
very large field of view $>$3\,sr compared to $<$0.03\,sr for the
IACTs, (ii) a duty cycle of $\sim$100\% compared to $<$10\% for
the IACTs. However, the EAS arrays generally suffer from poorer
angular resolution ($>$1$^{\circ}$ at 100\,TeV) as compared to the
IACTs ($<$0.1$^{\circ}$). The flux of $\gamma$-rays decreases
rapidly with energy and becomes extremely small at multi-TeV
energies thus, it becomes extremely difficult to detect it in
an overwhelming background of isotropic PCRs. Since the signal to
noise ratio for a telescope scales inversely with its angular
resolution thus, it becomes critically important to achieve as
small an angular resolution as possible which is precisely the
strategy devised for the GRAPES-3 EAS array.

So far 100\,TeV $\gamma$-rays have been detected from only a
handful of galactic sources including a supernova remnant and
a pulsar which might be sites of particle acceleration to PeV
energies within our galaxy \cite{Ahronian07,Amenomori19}. Most
of the sub-TeV $\gamma$-ray sources seem to be leptonic in
origin where $\gamma$-rays are produced via inverse Compton
scattering of synchrotron photons produced by TeV electrons.
In general, the inverse Compton scattering is strongly suppressed
by the Klein-Nishina effect at high energies and therefore, above
100\,TeV $\gamma$-rays are expected to be mostly of hadronic
origin. These $\gamma$-rays are produced during the decay of
neutral pions generated in the interactions PeV protons with the
ambient matter. Therefore, the observation of multi-TeV
$\gamma$-ray sources is essential for addressing the problem of
origin of PCRs \cite{Gaisser16}.

As discussed above, EAS arrays enjoy an advantage over the
IACTs in terms of longer observation time and a far greater
sky coverage and are therefore ideally suited for monitoring
a large number of sources and to search for new ones. But these
arrays suffer from poorer angular resolution thus making them
less competitive than the IACTs for the discovery of new sources. 
Typically, the IACTs have an excellent angular resolution of
5--8$^{\prime}$ in the sub-TeV energy region
\cite{Abdalla18,Aleksic16,Park15}. In EAS arrays the shower
direction is obtained from the information on the relative
arrival time of particles in the triggered detectors. These
particles are primarily electrons and muons which exhibit
considerable fluctuations due to various processes such as
multiple scattering and transverse momentum imparted during
the shower development. Therefore, the shower front acquires a
disc-like structure, about 1-2\,m thick near the central core
of the shower which increases with increasing distance from
the core. The mean energy of particles decreases away the core
due to delays caused by multiple scattering etc. Consequently, 
the shower front exhibits a curved profile away from the
shower core \cite{Bassi53}. A number of studies have shown the
the presence of a curvature in the shower front
\cite{Linsley61,Kozlov73,Hara83,Haeusler02}. Extensive Monte
Carlo simulation studies have also suggested the existence of
shower front curvature \cite{Piazzoli94,Battistoni98}. This
curvature needs to be corrected for accurate measurement of the
EAS direction and for that a precise measurement of the curvature
becomes an essential prerequisite for improving the angular
resolution of an EAS array. A discussion of the origin of shower
front curvature and finite disc structure can be found in a
monograph by Grieder \cite{Greider10}. The curved shower front
is adequately described by a cone centred on the shower axis for
distances up to 100--200\,m \cite{Acharya93,Ambrosio99}. By
implementing curvature correction an angular resolution of
1.1$^{\circ}$ at 150\,TeV was reported for the KGF EAS array
\cite{Acharya93}.

The EAS arrays record showers of varying size (N$_e$) and age (s).
While the size is proportional to energy of the PCR, the age
characterizes the stage of the shower development. In recent
past some attempts were made to study the relationship of the
shower front curvature on shower size
\cite{Acharya93,Agnetta97,Antoni01,Melcarne13}.
However, the work presented here is possibly the first
systematic investigation of the dependence of shower front
curvature on both the shower size and age. In \S\ref{time}
after a very brief overview of the GRAPES-3 array, the
methodology for measurement of the arrival times of EAS
particles is presented. In \S\ref{curve}, detailed
investigation of the shower front curvature is presented.
This is followed by \S\ref{angres} where the measurement
of the angular resolution by two distinctly different
techniques, namely, the odd-even and left-right methods are
presented. Next, a comparative study of the GRAPES-3 angular
resolution with other major arrays operating elsewhere in
the world is presented \S\ref{discuss}. Finally the conclusions
of this study are summarised in \S\ref{conclude}.

\begin{figure}[h]
\centering
\includegraphics[width=0.85\textwidth]{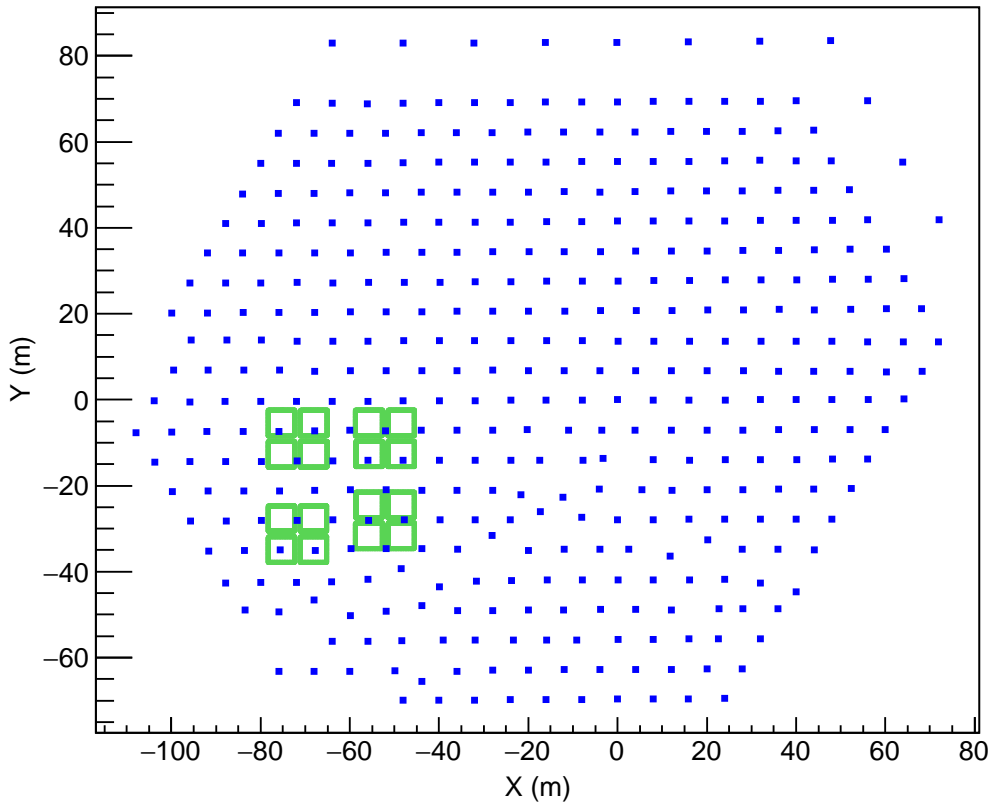}
\caption{\label{fig01} GRAPES-3 EAS array in Ooty, India
(11.4$^{\circ}$N, 76.7$^{\circ}$E, 2200m a.s.l). Small filled
squares represent scintillator detectors of 1\,m$^2$ area
each separated by 8\,m and the 16 big squares represent
muon modules of 35\,m$^{2}$ area each.} 
\end{figure}

\section{Measurement of time of shower particle}\label{time}
The GRAPES-3 experiment is located at Ooty, India at an altitude
of 2200\,m (11.4$^{\circ}$N, 76.7$^{\circ}$E). The two major components
of the array include (i) 400 plastic scintillator detectors (each
1\,m$^2$) spread over 25000\,m$^2$ \cite{Gupta05,Mohanty09} and a
tracking muon detector of 560\,m$^2$ area with a threshold of
1\,GeV \cite{Hayashi05}. A schematic of the array is shown in
Fig.\,\ref{fig01}. The scintillator array is triggered by EAS
produced by the PCRs in the energy range 10$^{12}$--10$^{16}$\,eV.
Over 10$^{9}$ EAS are recorded every year. A detailed description
of the trigger and the data acquisition may be found elsewhere
\cite{Gupta05}. Since the scintillator detectors are unshielded
they record particles that deposit a few MeV of energy which is
mainly contributed by the electrons, positron, muons and
$\gamma$-rays albeit with a rather small efficiency ($\sim$5\%).
Each scintillator detector is instrumented to measure both the
density and the arrival time of detected particles relative to
the trigger. This information is used to determine the energy
and the incident direction of the PCR responsible for producing
the EAS. 

Since the arrival direction of an EAS is determined by using
the relative arrival time of particles in the shower front
by the triggered detectors in the array, precise measurement
of arrival times is the key requirement for achieving a good
angular resolution. In GRAPES-3, the time is measured by a
32\,channel high performance time-to-digital converter (HPTDC)
which was developed in-house from an application specific
integrated circuit designed and developed by the Microelectronics
group at CERN, Geneva for the LHC experiments \cite{Gupta12}.

The signal from each scintillator is detected by a photomultiplier
tube (PMT) and transmitted through a 230\,m long, low-loss, co-axial
cable (5D2V) to the control room. Subsequently, the signals from
each scintillator is amplified, discriminated and its arrival
time relative to the EAS trigger is digitized by the HPTDC with
a resolution of 195\,ps. The HPTDCs utilize a quartz oscillator
to measure time, and therefore provide an exceptional performance
in terms of long-term stability and linearity especially compared
to the commercial TDCs.

The time T${_i}$ measured by detector ``i'' consists of two parts,
namely, T$_{Ei}$ the variable part contributed by particles in
the EAS, and second T$_{Zi}$ the passive part contributed by
the delay in PMT, co-axial cable, and signal processing electronics
etc. which is also called TDCZero. Ideally, T$_{Zi}$ should remain
constant, however, as shown later that this is not the case. To
achieve a good angular resolution, it is important to accurately
measure both of these parts. Although the signal from each
scintillator detector `i' to the HPTDC is transmitted by equal
length cables, however, T$_{Zi}$ shows significant variation from
detector to detector possibly due to the differences in the
response time of each detector, which includes the transit time
inside PMT, unequal propagation velocities in co-axial cables etc.

Traditionally, the T$_{Zi}$ for each detector in the GRAPES-3
experiment was measured relative to a common detector called
``muon paddle'' which is placed below the detector and used to
trigger on muons passing through the scintillator above. The
arrival time and charge contained in the PMT pulse produced
by through going muons is measured by HPTDC and a charge
integrating ADC, respectively for a duration of one hour.
The most probable value of the TDC distribution is defined as
the T$_{Zi}$ for detector `i'. By moving the muon paddle manually
from one detector to the next, the T$_{Zi}$ for every detector 
in the array is measured \cite{Gupta05}. One complete cycle of
measurements takes $\sim$40\,d owing to the fact that only
8--10 detectors can be manually calibrated in a single day.
However, it is observed that T$_{Zi}$ varies significantly
even on a short time scale of a day. This is subsequently
shown to be dependent on the ambient temperature. Thus, it
becomes essential to develop an alternative technique for
measuring T$_{Zi}$ on time scale shorter than a day. The
GRAPES-3 array records $\sim$10$^{5}$ EAS every hour and by
taking advantage of this high statistics data, an effective
technique has been developed to determine the T$_{Zi}$ on
hourly time-scale as explained below.

\begin{figure}[ht]
\vskip 0.1in
\centering
\includegraphics[width=0.75\textwidth]{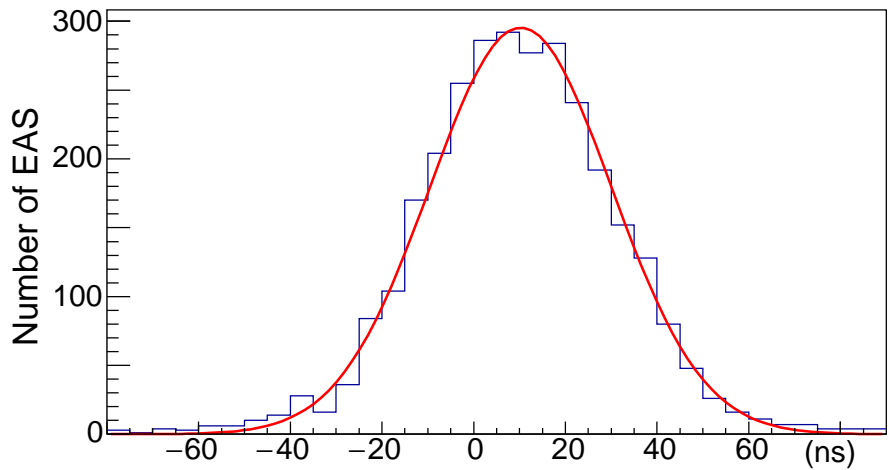}
	\caption{\label{fig02} Distribution of arrival time
                 difference of particles recorded by detector
                 1 and 9 for 1\,h. Gaussian fit to data yields
                 a peak at (10.3$\pm$0.4)\,ns and $\sigma$
                 of 19.9\,ns.}
\end{figure}

The distribution of arrival time difference ($\Delta$t) of any two
detectors (i,j) represents the distribution of projected angles
of the EAS incident in the array. The peak of $\Delta$t distribution
is a measure of the difference T$_{Zi}$-T$_{Zj}$. Fig.\,\ref{fig02}
shows the $\Delta$t distribution for detectors 9 and 24 (separated
by 16\,m) generated from one hour of EAS data. A Gaussian fit to
this distribution yields a peak value (10.3$\pm$0.4)\,ns and a
standard deviation $\sigma$=19.9\,ns. The value of $\sigma$ is
proportional to the detector separation. However, for accurate
direction reconstruction T$_{Z}$ needs to be measured for each
detector relative to a single reference detector in the array.
For detector further away from the reference detector, the EAS
would be of increasingly larger size and number of such EAS
would rapidly decline. In addition, the width $\sigma_{t}$ of the
$\Delta$t distribution increases proportional to the distance of
the detector and the reference detector. Therefore, estimating
T$_{Z}$ for distant detectors becomes rapidly inaccurate. For
example, for a distance of 80\,m $\sigma_{t}$=94\,ns. Also the
number of EAS reduces by more than a factor of two, thus, the
error in T$_{Z}$ increases to 2.4\,ns which is eight times larger
than for a distance of 16\,m. Below, a novel technique called
``random walk method'' which addresses this problem by proving
a robust, real-time and accurate estimate of the T$_{Z}$ for
every detector in the array relative to a common reference
detector is described.

\begin{figure}[ht]
\vskip 0.2in
\centering
\includegraphics[width=0.75\textwidth]{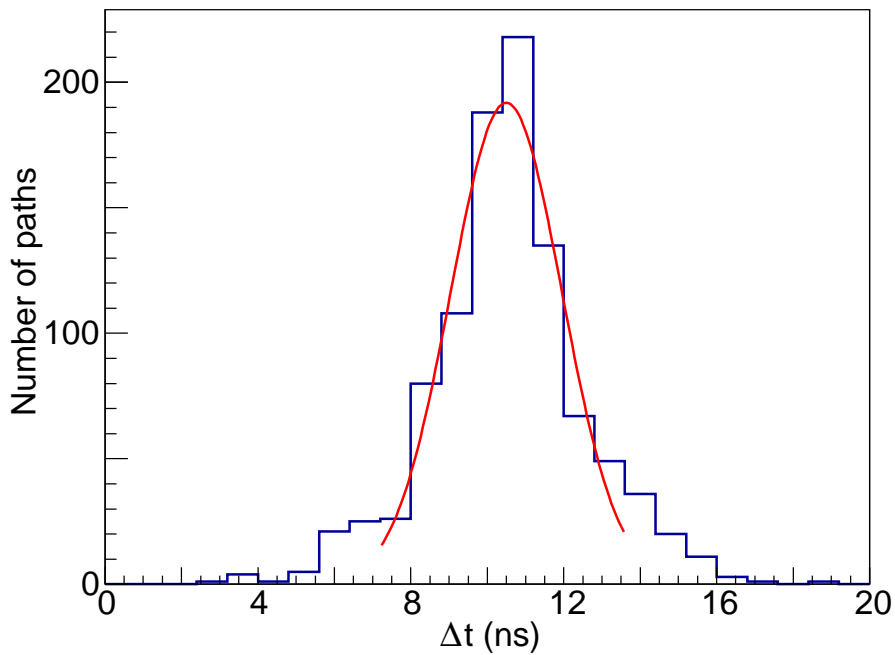}
	\caption{\label{fig03} Distribution of TDCZero `T$_{Z}$'
                 for detector 24 relative to reference detector
                 9 based on 1000 random walk paths. The peak
                 from a Gaussian fit occurs at (10.5$\pm$0.1)\,ns.}
\end{figure}

Here, the analysis of data collected during 2014 is presented.
During 2014, detector 9 had exhibited a very stable performance
and it had operated continuously without breakdown as well as
displayed a stable gain. Therefore, detector 9 is selected as
the reference
detector. As a first step, the peak value $\tau_{jk}$ (k=1,12)
of the $\Delta$t distribution of each detector `j' in the array
relative to its 12 neighbours is calculated. Here a pair of
detectors separated by 16\,m are termed ``neighbours''. Next,
starting from the reference detector 9 to any detector `i' can
be reached by a series of random steps to one of the 12 nearest
neighbours by generating a random number with 12 discrete values
which is equivalent to the rolling of a 12-sided dice. After the
first random step, only 11 out of the 12 random steps is allowed
to prevent the reverse step. This procedure is repeated until
the reference detector is reached after N steps. However, if N
exceeds 100, this random path is rejected and a new path is
generated. Then the sum t$_i$=$\Sigma\tau_{jk}$ for N steps is
calculated. A total of 1000 random paths are generated for the
EAS data of each hour and a distribution of t$_i$ is obtained
as shown in Fig.\,\ref{fig03} and the peak of this distribution
provides a measure of T$_{Zi}$. A Gaussian fit to this distribution
yields a value of (10.5$\pm$0.1)\,ns. The choice of limiting the
random path to 100 steps and total number of paths to 1000 
is based on the fact that increasing these values does not
result in any improvement in the accuracy of measured T$_{Zi}$.
However, when the detector performance is not stable the
the distribution of t$_i$ becomes broader and yet in most such
cases the peak of the distribution is unaffected and hence the
measured value of T$_{Zi}$ stays unchanged.

\begin{figure}[t]
\vskip 0.1in
\centering
\includegraphics[width=0.75\textwidth]{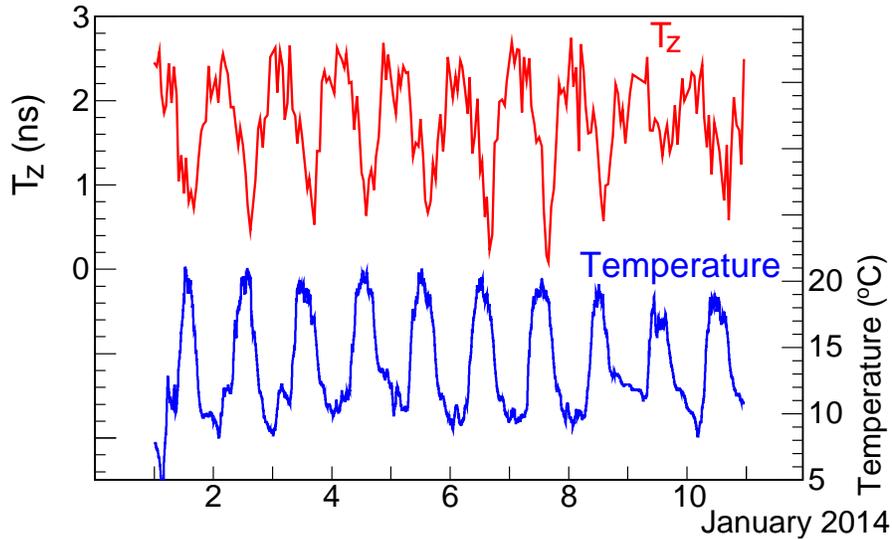}
	\caption{\label{fig04} Hourly variation of, (i) T$_{Z}$ for
                 detector 24, (ii) temperature during 1--10 January
                 2014. An anti-correlation of T$_{Z}$ and temperature
                 is observed.}
\end{figure}

In Fig.\,\ref{fig04} the hourly variation of T$_{Z}$ for detector 24
and the ambient temperature outside is shown during 1--10 January
2014. The variation of T$_{Z}$ is clearly anti-correlated with the
temperature and an excursion as large as 2.5\,ns in T$_{Z}$ is
observed in a single day which is significantly larger than the
0.195\,ns resolution of the HPTDC used to record time. The co-axial
cable used to transmit the PMT signal from detector 24 to the HPTDC
located in the control room was replaced and was lying exposed.
Thus, the variation of T$_{Z}$ is caused by the temperature
dependence of the dielectric constant of the insulator used in the
co-axial cable. Because, the detectors whose signal cable are buried
under the soil do not show such variation. However, the unique ability
to measure hourly T$_{Z}$ allows any variation in arrival time to be
accurately measured and corrected.

\section{Shower front curvature}\label{curve}
A major challenge in improving the angular resolution of an EAS
array depends on the ability to accurately measure the TDCZeros
in real time and then the shape of the shower front. It is known
that shower front does not have a plane shape and shows significant
curvature \cite{Oshima10}. Thus, the measurement of shower front
curvature becomes an important tool for improving the angular
resolution of an EAS array. But before measuring the curvature,
approximate direction of the EAS can be obtained by fitting a plane
to the observed arrival times of particles in each triggered
detector after subtracting respective T$_{Z}$ determined by the
random walk method described in \S\ref{time}. The plane shower
front is fitted by minimizing the quantity $\chi^2$

\begin{equation}\label{eq:plane}
	\rm \chi^2 = \Sigma_i(lx_i + my_i + nz_i + c(t_i^{obs}-t_0))^2
\end{equation}

Here, ${\rm l=sin\theta cos\phi}$, ${\rm m = sin\theta sin\phi}$
and ${\rm n = cos\theta}$ are the direction cosines of EAS axis
and ${\theta}$, ${\phi}$ are the zenith and azimuthal angles,
respectively. x$_{i}$, y$_{i}$, z$_{i}$ are the detector locations
and $\rm t_{i}^{obs}$ the arrival time measured by $\rm i^{th}$
detector. The shower front comprising of highly relativistic
particles moves at nearly the speed of light and therefore, the
velocity of light `c' is used as the velocity of the shower
front. Here, $\rm t_{0}$ is the reference time when the EAS hits
a fictitious detector located at the EAS core.

To determine the shower front curvature, the core location of the
EAS should be known. By fitting a lateral distribution function,
namely, the Nishimura-Kamata-Greisen (NKG) to the observed particle
densities by a maximum likelihood algorithm MINUIT, the core
location (X$_c$, Y$_c$) and other EAS parameters such as EAS size
`N$_e$' and age `s' are obtained \cite{Tanaka12}. The `N$_e$'
represents the total number of particles derived from the NKG fit
which mainly consist of electrons and a smaller fraction of muons.
About 98\% of EAS recorded by GRAPES-3 fall in the size range
10$^{3}$--10$^{6}$. The EAS age `s' represents the slope of the
lateral density of EAS particles and is a measure of the stage of
shower development. The value of s varies between 0 to 2 and with
1 representing the stage of maximum development. The average age
s\,=\,1.1 is obtained for the EAS recorded by GRAPES-3 experiment.
A detailed study of the dependence of shower front curvature on
both the shower size and age is described below.

The delay of the observed time relative to the expected arrival
time for a plane front for the i{\rm $^{th}$} detector is,

\begin{equation}\label{eq:planefit}
	\rm \delta t_i = t_i^{obs} - t^{plane}_i 
\end{equation}

where $\rm t^{plane}_{i}$ is given by,

\begin{equation}
{\rm t^{plane}_i = t_{0} - \frac{lx_{i} + my_{i} + nz_{i}}{c}} 
\end{equation}

Further, the distance of each detector $r_{i}$ from the EAS core
(X$_c$, Y$_c$) in the plane perpendicular to the EAS axis is
calculated as follow,

\begin{equation}
{\rm r_{i} = \sqrt{(x_i - X_c)^2 + (y_i - Y_c)^2 - D^2)}}
\end{equation}
 
where D is given by,

\begin{equation}
\rm D = (x_i-X_c) sin(\theta)cos(\phi) + (y_i-Y_c)sin(\theta)sin(\phi)
\end{equation}

\begin{figure}
\centering
\includegraphics[width=0.75\textwidth]{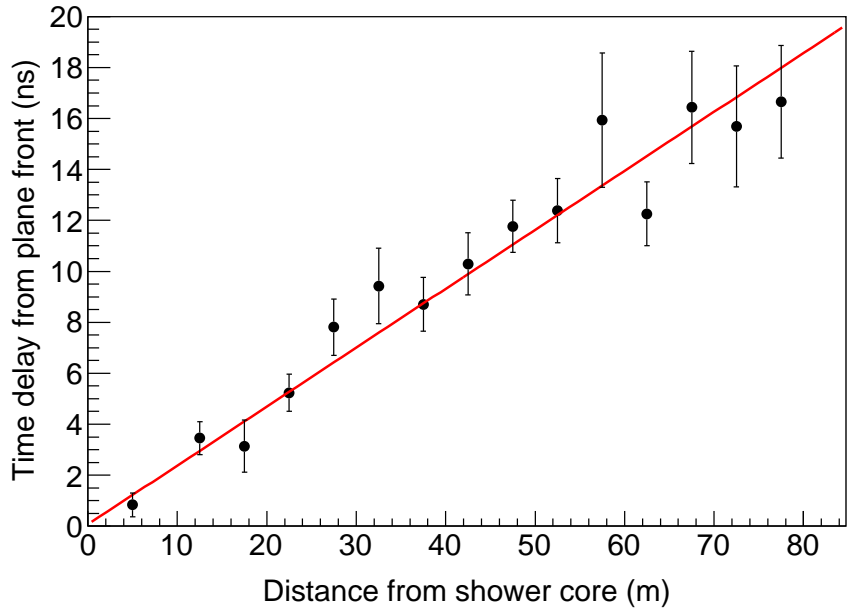}
	\caption{\label{fig05} Shower front curvature of an EAS
                 recorded on 1 January 2014. Number of triggered 
                 detectors is 261, size N$_{e}$\,=\,8.9$\times$10$^{4}$,
                 age s\,=\,1.21. A linear fit to the data represented
                 by line yields a slope of 0.23\,ns\,m$^{-1}$.}
\end{figure}

The profile of $\rm \delta t_{i}$ as a function of distance from the
core of an EAS is shown in Fig.\,\ref{fig05}. The linear dependence
of $\rm \delta t_{i}$ on distance from core is clear evidence of
curvature which is well described by a conical shape. From a linear
fit, the slope of the shower front `$\alpha$' is 0.23\,ns\,m$^{-1}$. 

\begin{figure}
\vskip 0.1in
\centering
\includegraphics[width=0.75\textwidth]{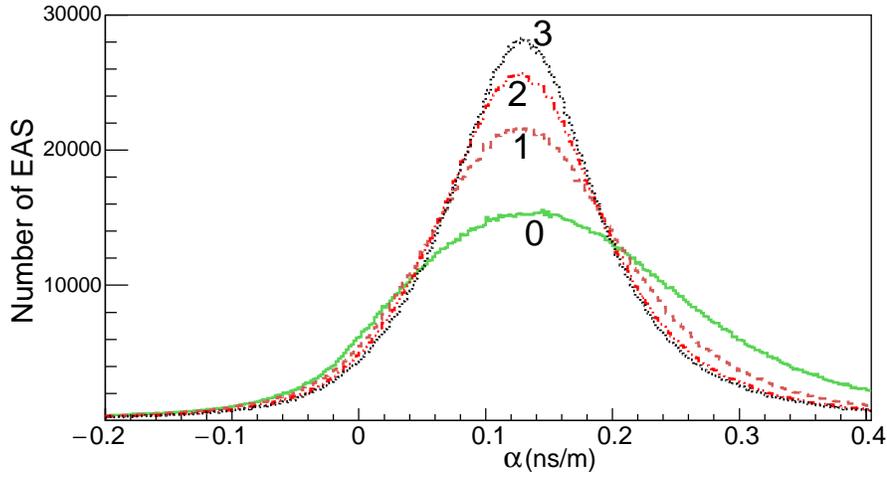}
	\caption{\label{fig06} Distribution of shower front slope
                 $\alpha$ from analysis of 2.4$\times$10$^{6}$ EAS.
                 (a) plot `0' represents slope distribution, (b)
                 `1' after removal of outliers, (c) `2' after
                 removal of second round of outliers, (d) `3'
                 after removal of third round of outliers.}
\end{figure}

$\alpha$ is determined for each EAS for the data collected during
1--10 January 2014. Cuts imposed on EAS for this study include,
$\rm \theta$$<$$40^{\circ}$, core within 30\,m from the center of
array. The distribution of $\alpha$ is shown by the plot labeled
`0' in Fig.\,\ref{fig06}. The mean and the root mean square
deviation (rms) of the $\alpha$ distribution are 0.189 and
0.305\,ns\,m$^{-1}$, respectively and its most probable value
$\rm \alpha_{peak}$ is 0.141\,ns\,m$^{-1}$. An examination of
individual showers shows the presence of outliers which can
significantly alter the fit and thereby influence the value of
$\alpha$. This could be a factor contributing to the large width
of $\alpha$. The outliers could be due to the noise in
electronics, or from unassociated particles / delayed hadrons in
the EAS etc. The following rigorous criteria are used to identify
and iteratively remove the outliers. Prima facie $\rm \delta t_{i}$
with large deviation from the expected conical shower front may
be treated as outliers. Since it is not possible to reliably
estimate $\alpha$ by fitting individual showers because of the
presence of outliers, $\rm \alpha_{peak}$ is used as the starting
estimate for every EAS. The outliers are identified by
calculating the time residuals for individual EAS by the
following expression,

\begin{equation}
\rm \tau_{res}^i = \delta t_{i} - \alpha_{peak} \times r_{i}
\end{equation}

Here, $\rm \alpha_{peak}$\,=\,0.141\,ns\,m$^{-1}$ and r$_{i}$ is
distance of the i{\rm $^{th}$} detector from EAS core. The mean
`$\mu$' and rms of residuals $\rm \tau_{res}^i$ are calculated 
for individual EAS. Thereafter, the data points with $\rm \tau_i$
such that $\rm \mu-(2 \times rms)$ $>\tau_i>$ $\mu+(2\times rms)$
are termed the ``outliers''. After removing outliers, each EAS
is again reconstructed by fitting a plane and slope $\alpha$
recalculated. The distribution of $\alpha$ after the removal of
outliers is shown by the plot labeled `1' in Fig.\,\ref{fig06}.
The mean value of $\alpha$ has decreased from 0.189 to
0.141\,ns\,m$^{-1}$ after the first iteration of outlier removal.
Moreover, the rms also decreased from 0.305 to 0.154\,ns\,m$^{-1}$.
This procedure for the removal of outliers and slope recalculation
is repeated two more times to obtain new slope values after each
iteration as shown in Fig.\,\ref{fig06} by the plots labeled `2',
`3' after the second and third iterations, repectively. After the
second and third iterations, the mean remained virtually
unchanged at 0.136\,ns\,m$^{-1}$ and the rms evetually reduced to
0.132\,ns\,m$^{-1}$. The fraction of detectors removed are 5\%,
5\% and 4\% after first, second and third iterations, respectively.
However, even after the third iteration, the rms is still large
and compareble to the mean indicating that the slope possibly
depends on other EAS parameters such as the size and age which
are inverstigated next.

\begin{figure}
\centering
\includegraphics[width=0.75\textwidth]{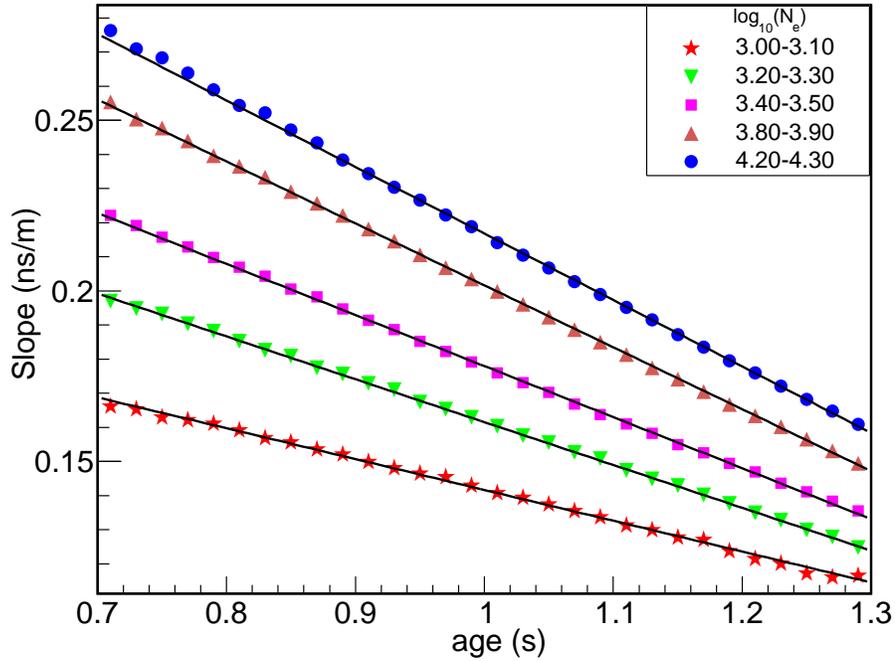}
	\caption{\label{fig07} Dependence of slope $\alpha$ on EAS
                 age (s) is shown for five size (N$_e$) groups.}
\end{figure}

For this analysis, the EAS recorded during the period from 1 January
to 31 December 2014 are used. The following selection cuts are
imposed on the data. Zenith angle ${\theta}$ is restricted below
40$^{\circ}$ and the EAS cores should fall within 30\,m from the
array centre. A total of 8$\times$10$^{7}$ EAS passed these cuts.
Further, the EAS are categorized into 20 logarithmic size groups
(each 10$^{0.1}$ wide) over the size range of 10$^{3}$--10$^{5}$.
The variations of EAS slope $\alpha$ as a function of age `s' for
the five out of 20 size groups are shown in Fig.\,\ref{fig07}. The 
values of $\alpha$ are obtained after the third iteration of
outlier removals. The following inferences can be drawn from
the plots shown in Fig.\,\ref{fig07}, (i) for any given shower
size the slope $\alpha$ systematically decreases with increasing
age, (ii) for a given shower age, $\alpha$ is larger for higher
shower size.

A linear fit provides an excellent description for the dependence
of slope $\alpha$ on age (s) for each of the size group as seen
from Fig.\,\ref{fig07}. From the linear fits to the data of 20
size groups the slope `M' and intercept `C' parameters are obtained
for each case. Thus, the slope $\rm \alpha_{s}^{N_{e}}$ may be
parameterized as,

\begin{equation}
\rm \alpha_{s}^{N_{e}} = M_{N_{e}} \times s + C_{N_{e}}
\end{equation}

\noindent where $\rm M_{N_{e}}$ and $\rm C_{N_{e}}$ are the slope and
intercept, respectively from the linear fit shown in
Fig.\,\ref{fig07}. 

\begin{figure}[ht]
\centering
\includegraphics[width=0.75\textwidth]{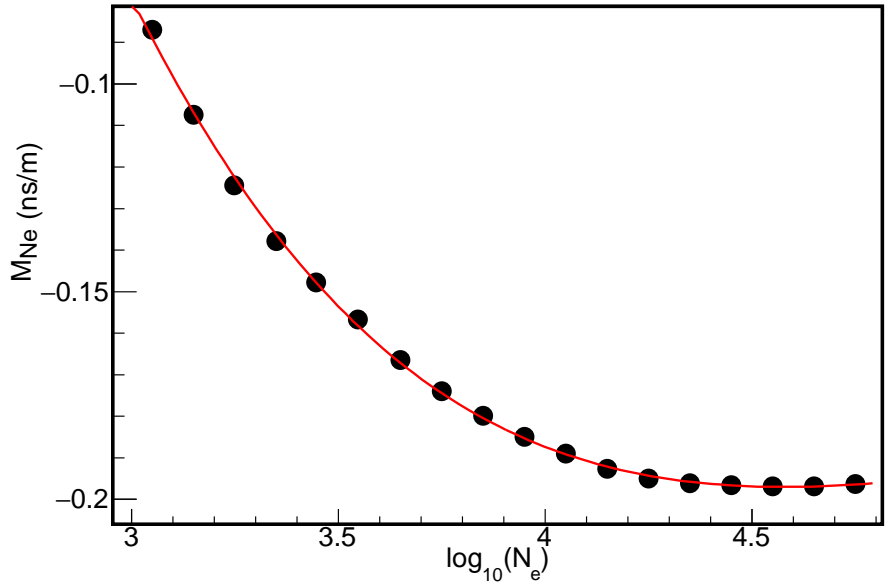}
	\caption{\label{fig08} Dependence of $\rm M_{N_{e}}$
                 on shower size. $\rm M_{N_{e}}$ is the slope
                 in the linear fit to the data displayed in
                 Fig.\,\ref{fig07}.}
\end{figure}

\begin{figure}[ht]
\centering
\includegraphics[width=0.75\textwidth]{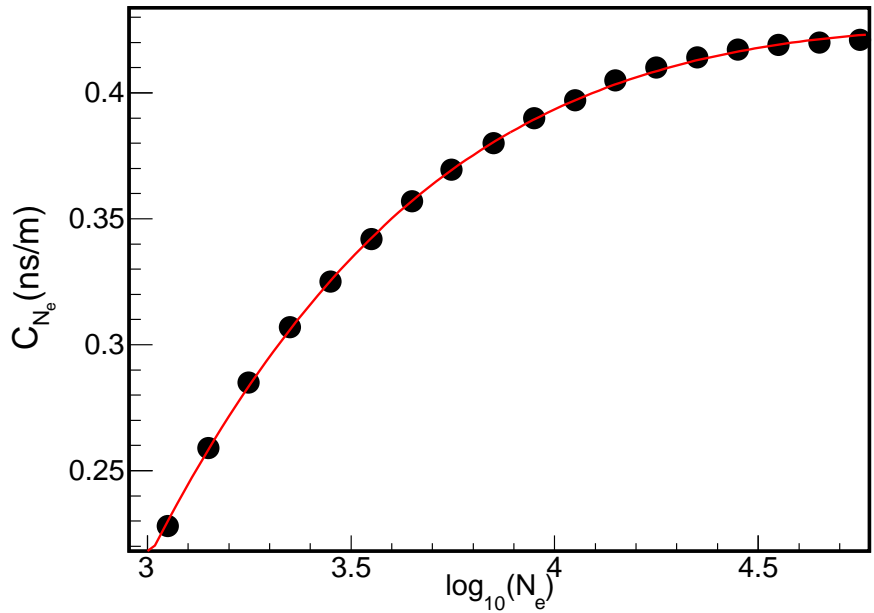}
	\caption{\label{fig09} Dependence of $\rm C_{N_{e}}$ on shower
                 size. $\rm C_{N_{e}}$ is intercept in the linear fit
                 to data shown in Fig.\,\ref{fig07}.}
\end{figure}

Here, the parameter $\rm M_{N_{e}}$ and $\rm C_{N_{e}}$ represents
the slope and intercept, repectively of the linear fit. In
Fig.\,\ref{fig08} the variation of parameter $\rm M_{N_{e}}$ with
shower size is shown. The magnitude of $\rm M_{N_{e}}$ increases 
with increasing N$_e$, which indicates that the rapidity with
which the slope $\alpha$ increases with age `s' also increases
with size N$_e$. Since the dependence of $\rm M_{N_{e}}$ on N$_e$
apears to be somewhat similar to an exponential, an exponential
fit was tried. However, these data required the use of two
exponentials of the form,
$\rm a_1\,exp(-b_1\,log_{10}N_e) + c_1\,exp(-d_1\,log_{10}N_e^2)-1$.
The values of these parameters derived from the fit are,
a$_1$\,=\,0.475\,ns/m, b$_1$\,=\,-0.036, c$_1$\,=\,1.792\,ns/m
and d$_1$\,=\,0.216. The fit to $\rm M_{N_{e}}$ obtained by using
these parameters displayed in Fig.\,\ref{fig08} shows excellent
agreement between the data and the fit.

Similarly, the variation of intercept $\rm C_{N_{e}}$ as a function
of N$_e$ is shown in Fig.\,\ref{fig09}. This dependence also
appears to be similar to an exponential and could be fitted well
by a combination of two exponential functions given by
$\rm 1-a_2\,exp(-b_2\,log_{10}N_e) - c_2\,exp(-d_2\,(log_{10}N_e)^2)$.
The parameters from this fit are, a$_2$\,=\,0.528\,ns/m,
b$_2$\,=\,-0.081, c$_2$\,=\,1.014\,ns/m, d$_2$\,=\,0.159. This
analysis clearly shows that the shower front curvature is not a
constant but varies from shower to shower and yet can be uniquely
estimated from the knowledge of EAS size and age. The main
objective of this study is to investigate if such a
parametrization of shower front curvature could actually lead to
significant improvement in the angular resolution of the array.
The impact of shower dependent curvature correction is
qunatitatively examined in the next section. 

\section{Angular resolution of GRAPES-3 array}\label{angres}

\begin{figure}
\centering
\includegraphics[width=0.75\textwidth]{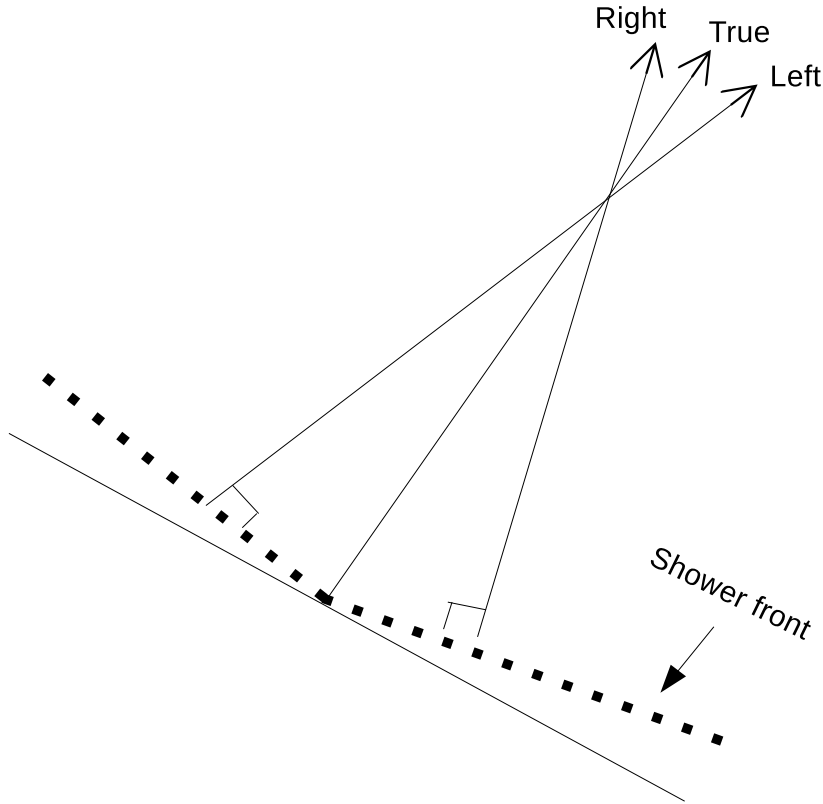}
	\caption{\label{fig10} A schematic representing the
	         biased direction due to the presence of shower
                 front curvature with left-right method.}
\end{figure}

In \S\ref{curve}, a detailed study of the shower front curvature
showed that its dependence on the EAS size and age can be easily
parametrized. Here, the EAS direction is corrected for shower
front curvature and the effect of this correction on the angular
resolution of the array is investigated. For this purpose two
independent techniques of dividing the array into, (i) the ``odd''
and ``even'' numbered detectors, (ii) a ``left'' and a ``right''
sub-arrays are implemented. The details of these two techniques
are described in an earlier work from GRAPES-3 experiment
\cite{Oshima10}.

The scintillator detectors in the GRAPES-3 array are deployed
with a hexagonal geometry as shown in Fig.\,\ref{fig01}. These
detectors are labeled sequentially, starting from the centre
of the array and thereafter proceeding clockwise along the
first hexagonal ring, followed by the second ring and so on.
For the odd-even method, the GRAPES-3 array is divided into
two sub-arrays. Each sub-array comprises of either ``odd'' or
``even'' numbered detectors. By using the plane shower front
independent estimate of the arrival direction is obtained
from the ``odd'' ($\theta_o$,$\phi_o$) and ``even''
($\theta_e$,$\phi_e$) sub-arrays. Next, the space angle
$\psi_{oe}$ between these two directions is calculated. The
distribution of $\psi_{oe}$ provides a measure of the 
angular resolution of the array by the odd-even method. 
For the left-right method, the array is divided into a ``left''
and a ``right'' sub-array by the following criterion. For each
shower, a line joining its core and the array center is used
as the dividing line for the two sub-arrays. Therefore, unlike
the odd-even method, the detectors in the ``left'' and ``right''
sub-arrays change with every shower according to the location 
of the core.

The space angle $\psi_{lr}$ between the two directions measured
by the ``left'' ($\theta_l$, $\phi_l$) and ``right'' ($\theta_r$,
$\phi_r$) sub-arrays would be used for further analysis. The
distribution of $\psi_{lr}$ can not be used as a measure of the
angular resolution because of the contribution from the shower
front curvature. This curvature results in a systematic tilt of
the reconstructed ``left'' and ``right'' directions as shown
schematically in Fig.\,\ref{fig10} for a conical shower front.
This systematic tilt between ``left'' and ``right'' directions
can be used to estimate the shower front curvature on a event
by event basis. On the other hand, the odd-even method is not
sensitive to the existence of shower front curvature. This is
because the ``odd'' and ``even'' sub-arrays are basically two
overlapping arrays each with half the detectors. Thus, for a
curved
shower front both the sub-arrays provide directions which may
be inaccurate, but both point in the same incorrect direction
within the angular resolution of the arrays. In summary, the
angle $\psi_{oe}$ provides an estimate of the best achievable
angular resolution because it only contains the statistical
errors since the systematic errors are eliminated because of
the overlapping nature of the two sub-arrays. Consequently,
odd-even method can not provide the absolute direction of an
EAS. On the other hand, if appropriate correction can be made
for the shower front curvature then the left-right method
provides the correct and absolute direction of an EAS relative
to the local frame of reference. 

For the left-right study, the direction of each EAS ($\theta$,
$\phi$) is obtained by fitting a plane shower front as given
in Eq.~\ref{eq:plane}. Next, by using the size and age of the
EAS as described in \S\ref{curve}, the slope of shower front
curvature is calculated by assuming it to be cone shaped and
corrected as follows,

\begin{equation}
\rm t_{c}^{i} = t_{m}^{i} - r_i \times \alpha_{s}^{N_{e}}
\end{equation}

Here $\rm t_{m}^{i}$ are measured arrival times, ${\rm r_i}$
distances of the i{\rm $^{th}$} detector from the shower core,
${\rm \alpha_{s}^{N_{e}}}$ the size and age dependent slope of
the shower curvature as described in \S\ref{curve}. Therefore,
the arrival times $\rm t_{c}^{i}$ obtained after the removal
conical shower front curvature, effectively represents a plane
shower front. Therefore, the EAS direction ($\theta$, $\phi$)
can be obtained by a plane fit. Following the plane fit the
outliers in the data are iteratively removed as described in
\S\ref{curve}.

\begin{figure}
\centering
\includegraphics[width=0.75\textwidth]{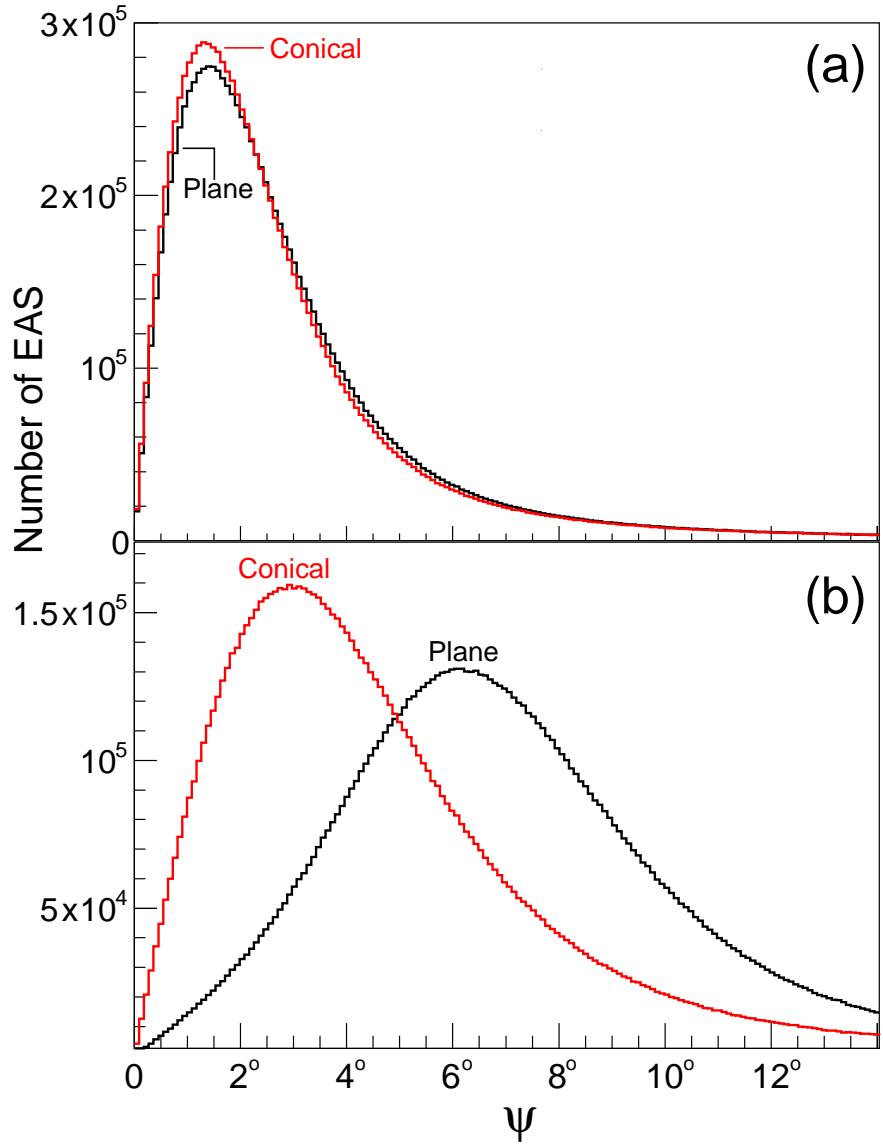}
	\caption{\label{fig11} Distribution of space angles $\psi$,
                 between two sub-arrays for, (a) odd-even for plane
                 and conical fits, (b) left-right also for plane
                 and conical fits for shower size $>$10$^{4.0}$.}
\end{figure}

The distribution of space angle between the ``odd'' and ``even''
sub-arrays for the shower size $\geq$10$^{4.0}$ for plane and
conical shower fronts are shown in Fig.\,\ref{fig11}a. These
the two distributions appear indistingushable which is not
surprising the two sub-arrays are nearly identical and fully
overlap and therefore, the EAS directions measured by them
contain identical systematic effects which get eliminated
when the space angle between them is measured. Thus, the
width of the distribution in Fig.\,\ref{fig11}a reflects 
the true angular resolution of the sub-arrays. 

However, when this procedure is repeated for the ``left''
and ``right'' sub-arrays, the outcome is completely different
as shown in Fig.\,\ref{fig11}b. The distribution for the
Plane front fit is almost a factor of two wider than for the
conical shower front as seen in  Fig.\,\ref{fig11}b. This
clearly shows that the shower curvature plays a significant in
direction reconstruction and the curvature correction can
improve the angular resolution of the array.

\begin{figure}
\centering
\includegraphics[width=0.75\textwidth]{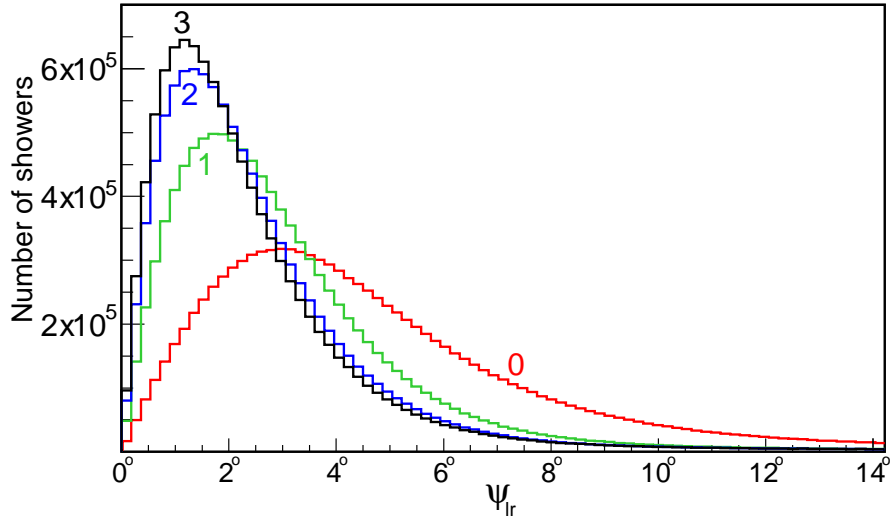}
	\caption{\label{fig12} Distribution of space angle
                 $\rm \psi_{lr}$ by left-right method for
                 N$_{e}\,>\,$10$^{4}$ after iterative outlier
                 removals, (i) `0' without outlier removal,
                 (ii) `1' first, (iii) `2' second, (iv) `3'
                 third iteration of outlier removals,
                 respectively.}
\end{figure}

As discussed earlier in \S\ref{curve}, the outliers in the data
significantly degrade the measurement of arrival direction of
an EAS by the left-right method, thereby influencing the
achievable angular resolution of the array. To improve the
resolution, direction reconstruction needs to be carried out
after iterative removal of the outliers as described in
\S\ref{curve} and the resulting histograms are shown in
Fig.\,\ref{fig12} for shower size N$_e\,>\,$10$^4$. The width of
each histogram is representative of the angular resolution
achieved. Histograms labeled, (i) `0' is without outlier removal,
(ii) `1' after the first round of outlier removal, (iii) `2'
after the second round of outlier removal, (iv) `3' after the
third round of outlier removal. A reduction in the width of the
histogram indicating corresponding improvement in the angular
resolution of the array is observed after each round of outlier
removal. The distribution becomes significantly narrower after
the first round of outlier removal and the improvement is more
modest after the second and third round. No measureable reduction
in the histogram width is observed after the fourth and higher
rounds of outlier removal. Consequently, only three iterations
of outlier removal are carried out hereafter.

\begin{figure}[t]
\vskip 0.0in
\centering
\includegraphics[width=0.85\textwidth]{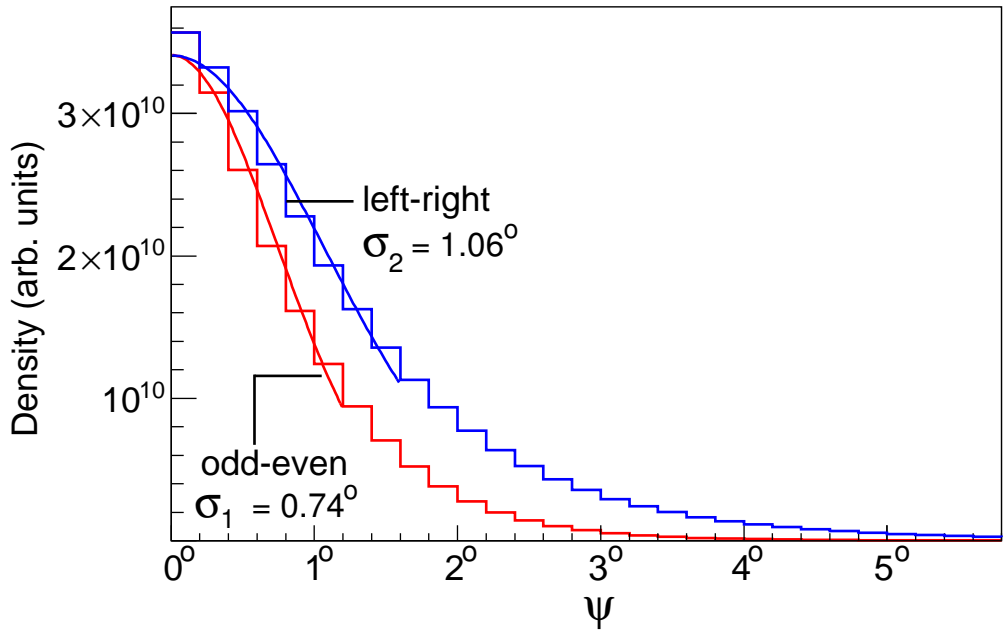}
	\caption{\label{fig13} Event density per unit solid
                 angle as a function of space angle $\psi$ for
                 N$_e>$10$^{4.0}$ for the left-right sub-arrays.}
\end{figure}

The distribution of space angle $\rm \psi$ obtained from the two
sub-arrays is converted into a density distribution by dividing
the contents of each angular bin by its solid angle.
If $\rm \psi_{low}$ and $\rm \psi_{upper}$ are the lower- and upper
boundaries of an angular bin, then its solid angle is,
cos$\rm \psi_{low}$\,--\,cos$\rm \psi_{upper}$. The two density
distributions as a function of space angles $\rm \psi$ for the
odd-even and left-right sub-arrays, respectively are shown in
Fig.\,\ref{fig13} for shower size N{\rm $_e\,\geq\,10^{4.0}$}.
The angular resolution of an array can be defined in a variety
of ways. 
For a Gaussian density distribution, it can be shown that the
maximum signal to noise (S/N) ratio for a point source occurs
for an opening space angle $\rm \psi_0$\,=\,1.6$\times\sigma$,
where $\sigma$ is the standard deviation of corresponding
density distribution. Alternatively, for a point $\gamma$-ray
source the maximum S/N ratio after the reconstruction of
showers would occur for a space angle of $\rm \psi_0$ centred
on the source direction. This relation is used subsequently
to estimate the angular resolution $\sigma$ of the GRAPES-3
array given by $\sigma$\,=\,$\rm \psi_0$/1.6
\cite{Acharya93,Graham94}. For the odd-even method, each
sub-array covers nearly the full area of array, except that
(i) the number of detectors are halved, (ii) the space angle
$\rm \psi$ contains errors from measurements by each of the
two sub-arrays. These two effects increase the errors by a
factor of $\sqrt{2}$ each, and therefore, the true error
$\rm \sigma_{oe}$ is a factor of $\sqrt{2}\times\sqrt{2}$\,=\,2
smaller than the measured error $\sigma_1$ obtained from a 
Gaussian fit to the histogram labeled `odd-even' in
Fig.\,\ref{fig13}. On the other hand, for the left-right
method, apart from the two effects mentioned above, there is
a third effect due to the decrease in the area of each
sub-array by a factor of two. Thus, the true error
$\rm \sigma_{lr}$ is a factor of 2$\sqrt{2}$ smaller than the
measured $\sigma_2$ obtained from a Gaussian fit to the
histogram labeled `left-right' in Fig.\,\ref{fig13} and thus,
\begin{equation}
\rm \sigma_{1}\,=\,\sqrt{2}\times\sigma_{2}
\end{equation}
As discussed above, the density distribution for the left-right
method shown in Fig.\,\ref{fig13} should be broader than the
odd-even method by a factor $\sqrt{2}$\,=\,1.41. The Gaussian
fits to the two density distributions shown in Fig.\,\ref{fig13}
yield, $\rm \sigma_{1}$\,=\,0.74$^{\circ}$ and
$\rm \sigma_{2}\,=\,1.06^{\circ}$ which yields a ratio of 1.43
which is close to the expected value of 1.41.

\begin{figure}
\centering
\includegraphics[width=0.75\textwidth]{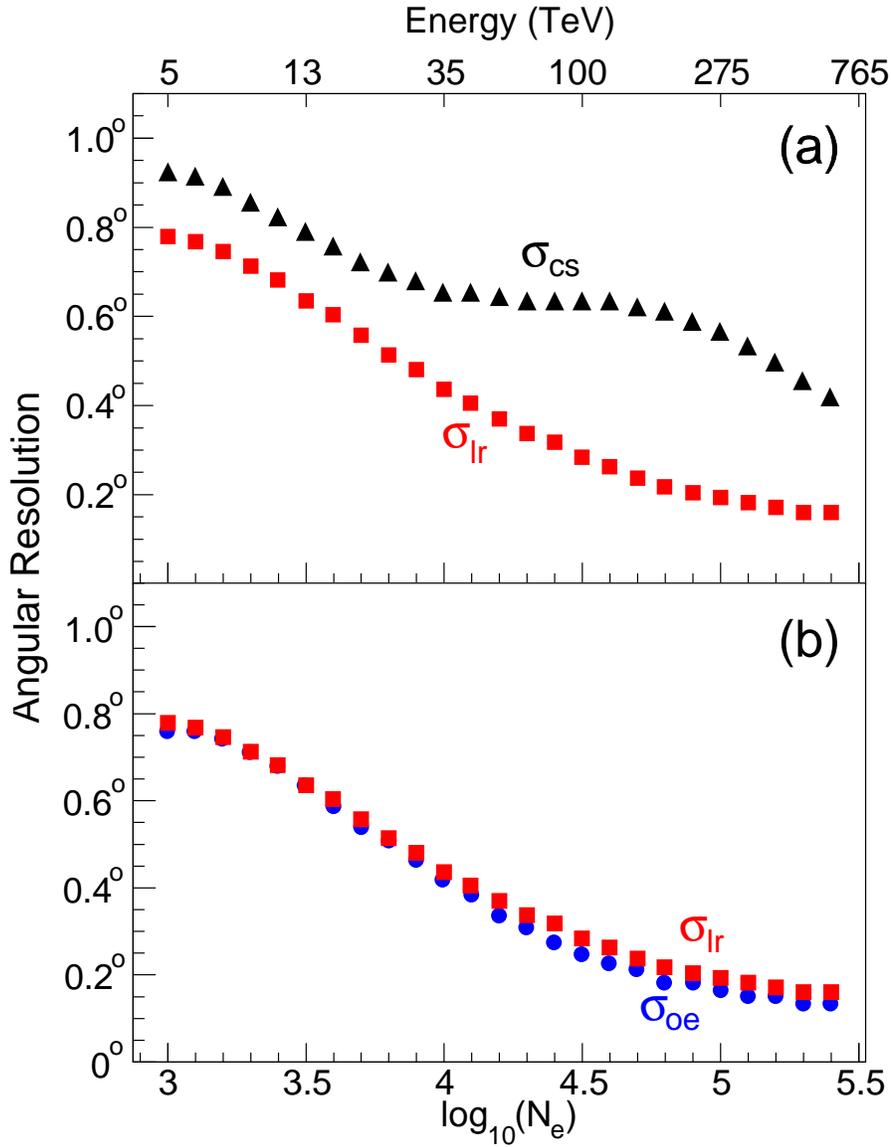}
	\caption{\label{fig14} Variation of angular resolution as
                 a function of shower size N$_{e}$ for (a) a
                 constant shower front curvature
                 determined with size and age dependent slope
                 and fixed slope for left-right method as a
                 function of integral shower size.}  
\end{figure}

In Fig.\,\ref{fig14}a the variation of angular resolution
$\sigma_{cs}$ of the GRAPES-3 array obtained by using a constant
slope of 0.129\,ns\,m$^{-1}$ for the shower front curvature,
after three iterations of outlier removal is shown. The angular
resolution $\sigma_{cs}$ decreases from 0.92$^{\circ}$ (55$^{\prime}$)
for size N$\rm _{e} > 10^{3.0}$ to 0.42$^{\circ}$ (25$^{\prime}$)
for N$\rm _{e} > 10^{5.5}$. Next, the variation of $\sigma_{lr}$
after the corrections for the size and age dependent shower
front curvature and three rounds of iterative removal of
outliers is shown in Fig.\,\ref{fig14}a. $\sigma_{lr}$ decreases
from 0.8$^{\circ}$ (47$^{\prime}$) to just below 0.16$^{\circ}$
(10$^{\prime}$) over the same shower size range. Clearly the
angular resolution measured after the size and age dependent
correction is significantly better than the constant slope
correction generally used. The improvement is especially
significant for higher shower sizes.

\begin{figure}
\centering
\includegraphics[width=0.85\textwidth]{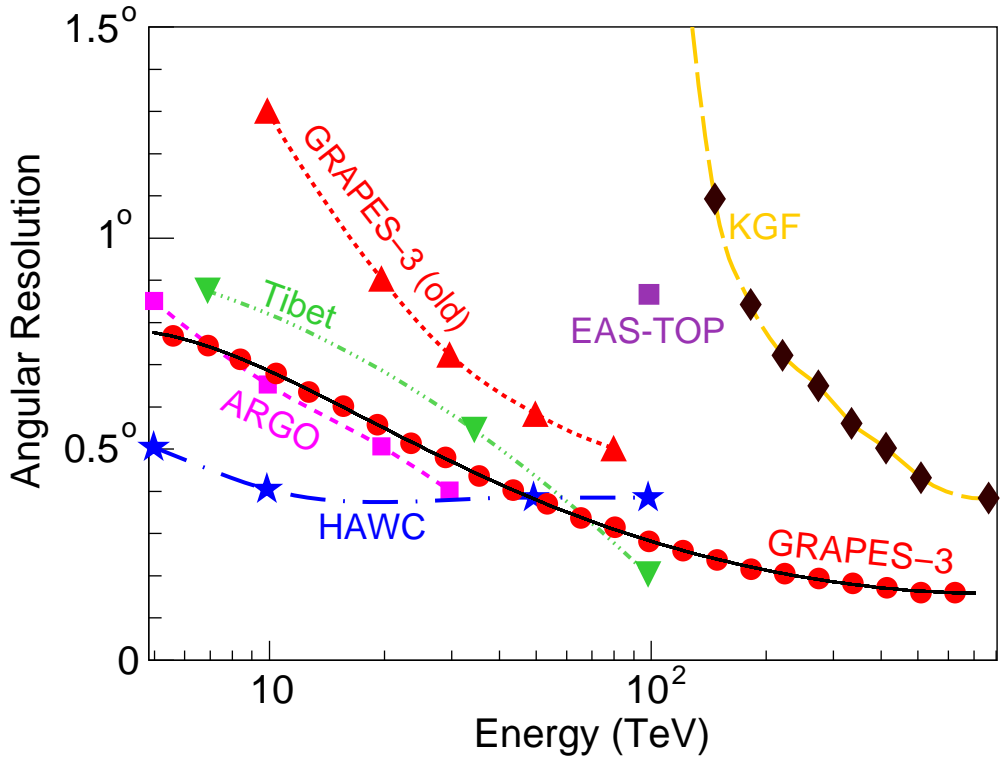}
	\caption{\label{fig15} Variation of angular resolution
                 with PCR energy for following arrays, (1) KGF
                 100--800\,TeV \cite{Acharya93}, (2) EAS-TOP
                 100\,TeV \cite{Aglietta91}, (3) GRAPES-3
                 10--80\,TeV old analysis \cite{Oshima10}, (4)
                 Tibet AS$\gamma$ 7--100\,TeV \cite{Amenomori93},
                 (5) ARGO-YBJ 5--30\,TeV \cite{Aielli12}, (6)
                 HAWC 5--100\,TeV \cite{Alfaro17}, (7) GRAPES-3
                 5--800\,TeV present analysis.}
\end{figure}

As mentioned earlier, the odd-even method provides the best
achievable angular resolution because this method eliminates 
common systematic errors and only the statistical uncertainties
in the data survive. In Fig.\,\ref{fig14}b the angular
resolution $\rm \sigma_{oe}$ obtained by the this method is
shown as a function of the shower size. $\rm \sigma_{oe}$ varies
from 0.76$^{\circ}$ (46$^{\prime}$) to 0.13$^{\circ}$ (8$^{\prime}$)
over the same size range. The angular resolution $\rm \sigma_{lr}$
obtained by the left-right method as a function of shower size
is overlaid in Fig.\,\ref{fig14}b. The two methods display very
similar bahviour, clearly indicating that the use of real time
TDCZeros as well as the size and age dependent shower front
curvature correction resulted in not only a sizable improvement
in the angular resolution, but also the systematic effects are
almost completely eliminated. Thus, the angular resolution is
now dominated by the statistical uncertainties present in the
GRAPES-3 data obtained by the odd-even method.

\section{Discussions}\label{discuss}
Worldwide, a number of experiments have been operated to detect
multi-TeV $\gamma$-ray sources. The experiments that are no
longer collecting data include the KGF, the EAS-TOP and the
ARGO-YBJ arrays. Located at an atmospheric depth of
920\,g.cm$^{-2}$ the angular reolution of KGF array as a function
of PCR energy displayed by symbol $\blacklozenge$ in
Fig.\,\ref{fig15} varies from 1.1$^{\circ}$ at $>$150\,TeV to
0.4$^{\circ}$ at $>$800\,TeV \cite{Acharya93}. The dashed line
joining the data is not a fit but meant to guide the eye. The
EAS-TOP array (atmospheric depth 820\,g.cm$^{-2}$) had an angular
resolution of 0.9$^{\circ}$ at $>$100\,TeV shown by symbol
$\blacksquare$ in Fig.\,\ref{fig15} \cite{Aglietta91}. The
ARGO-YBJ (depth 600\,g.cm$^{-2}$), a sophisticated RPC based
carpet array had an angular resolution that varied from
0.9$^{\circ}$ at $>$5\,TeV to 0.4$^{\circ}$ at $>$30\,TeV shown by
the symbol $\blacksquare$ in Fig.\,\ref{fig15} \cite{Aielli12}.

The EAS arrays operating currently include the Tibet AS$\gamma$,
the HAWC water Cherenkov and of course the GRAPES-3. The
angular resolution of Tibet AS$\gamma$ array (depth
600\,g.cm$^{-2}$) varies from 0.9$^{\circ}$ at $>$7\,TeV to
0.2$^{\circ}$ at $>$100\,TeV shown by symbol $\blacktriangledown$ in
Fig.\,\ref{fig15} \cite{Amenomori93}. Interestingly, the angular
resolution of HAWC array (depth 600\,g.cm$^{-2}$) varies within a
narrow range of 0.5$^{\circ}$--0.4$^{\circ}$ in the energy range
$>$5--100\,TeV as shown by symbol $\bigstar$ in Fig.\,\ref{fig15}
\cite{Alfaro17}. The angular resolution of GRAPES-3 (depth
800\,g.cm$^{-2}$) by using the old method varied from 1.3$^{\circ}$
at $>$10\,TeV to 0.5$^{\circ}$ at $>$80\,TeV as shown by symbol
$\blacktriangle$ in Fig.\,\ref{fig15}. But the new real time 
TDCZero as well as size and age dependent shower front curvature
correction led to a significant improvement in the resolution,
compared to the previous analysis \cite{Oshima10}. The angular
resolution improved from 1.3$^{\circ}$ to 0.7$^{\circ}$ at $>$10\,TeV
and from 0.5$^{\circ}$ to 0.3$^{\circ}$ at $>$80\,TeV in the new
analysis as shown by symbol ${\bullet}$ in Fig.\,\ref{fig15}. In
view of the sub-degree values, hereafter the angular resolution
of the GRAPES-3 shall be quoted in arc-minutes. 

But even more significantly this analysis has permitted the
reconstruction of showers of energies as low as $>$5\,TeV with a
respectable resolution of 47$^{\prime}$ by the GRAPES-3 array
located deep in the atmospheric at a depth of 800\,g.cm$^{-2}$.
This might prove to be a critical factor in the detection of
TeV $\gamma$-ray sources, most of which show steepening of
energy spectrum in the TeV region. The EAS-TOP array too had
operated at an atmospheric depth similar to the GRAPES-3, yet
the GRAPES-3 angular resolution is a factor of three better
\cite{Aglietta91} as shown in Fig.\,\ref{fig15}. The angular
resolution of GRAPES-3 is comparable to that of ARGO-YBJ
\cite{Aielli12} and Tibet AS$\gamma$ \cite{Amenomori93} as
shown in Fig.\,\ref{fig15}, despite the fact that both
these arrays are located at 200\,g.cm$^{-2}$ shallower depth
than GRAPES-3. Only the large HAWC array \cite{Alfaro17}
also located 200\,g.cm$^{-2}$ shallower than GRAPES-3 provides
a better resolution than the GRAPES-3 in 5--50\,TeV region.
But above 50\,TeV, GRAPES-3 delivers a slightly better
performance. The key takeaways from this analysis which is
based on the use of real time TDCZeros as well as size and
age dependent shower front curvature corrections are, (i)
For similar atmospheric depths, the method described here
yields a better angular resolution, (ii) The improvement in
angular resolution is equivalent to arrays operating at
200\,g.cm$^{-2}$ shallower depths. 

\section{Conclusions}\label{conclude}
A new method has been developed to determine the real time delay
caused by the photomultiplier tubes, co-axial cables, and
electronics in each detector of the array by using a statistical
approach. This is followed by an accurate modeling of the shower
front curvature using the data collected by the GRAPES-3 array
during 2014. This data showed that shower front curvature can be
modeled by a conical shape which depends on the shower size
and age. This dependence is accurately parametrized in the
GRAPES-3 data. The shower front curvature shows clear dependence
both on the shower size and age. The correction for this
curvature allowed the GRAPES-3 array to achieve an angular
resolution which is dictated primarily by the uncertainties in
the data. The angular resolution varies from 47$^\prime$ at
$>$5\,TeV to 10$^\prime$ at $>$500\,TeV. These values compare
favourably with the arrays such as ARGO-YBJ, Tibet AS$\gamma$
and HAWC located at higher elevations than GRAPES-3 where the
atmospheric depth is 200\,g.cm$^{-2}$ shallower. This method if
applied to existing arrays might lead to improvement in their
angular resolution and thereby enhance their sensitivity for
the discovery of fainter $\gamma$-ray sources.

\ack 
We dedicate this work to the memory of Prof. B.V. Sreekantan who
had guided and mentored several of us and who passed away recently
in Bangalore, India. We thank D.B. Arjunan,  V. Jeyakumar, S.
Kingston, K. Manjunath, S. Murugapandian, S. Pandurangan, B. Rajesh, K. Ramadass,
V. Santoshkumar, M.S. Shareef, C. Shobana, R. Sureshkumar for their 
contributions in the efficient running of the experiment.

\end{document}